\documentclass[prd,nofootinbib,preprint,superscriptaddress]{revtex4-1}

\usepackage{amsmath, amssymb, amsthm, graphicx, epsfig, fancyhdr,epsfig}

\usepackage{tikz-feynman}
\tikzfeynmanset{compat=1.1.0}

\usepackage{tikzsymbols}

\usepackage{float}
\usepackage{xcolor}

\newcommand{\be}{\begin{equation}}
\newcommand{\ee}{\end{equation}}
\newcommand{\bea}{\begin{eqnarray}}
\newcommand{\eea}{\end{eqnarray}}

\newcommand{\ag}[1]{\textcolor{blue}{[Anish: #1]}}

\begin{document}

\title{Radiative Plateau Inflation with Conformal Invariance: \\ \it{Dynamical Generation of Electroweak and Seesaw Scales} }

\author{Anish Ghoshal}
\email{anish.ghoshal@fuw.edu.pl}
\affiliation{\,\,Institute of Theoretical Physics, Faculty of Physics, University of Warsaw,\\ ul. Pasteura 5, 02-093 Warsaw, Poland }

\author{Nobuchika Okada}
\email{okadan@ua.edu}
\affiliation{Department  of  Physics  and  Astronomy, \\ University  of  Alabama,  Tuscaloosa,  Alabama  35487,  USA}

\author{Arnab Paul}
\email{arnabpaul9292@gmail.com  }
\affiliation{Indian Statistical Institute, \\ 203, B.T. Road, Kolkata-700108, India}

\begin{abstract}
\it{We investigate a scale-invariant $B-L$ scenario where the Standard Model (SM) is supplemented with a dark scalar $\phi$ which has gauge \& Yukawa interactions, with the couplings $g_{BL}$ and $y$, respectively, leading to radiative plateau inflation at scale $\phi=M$ in the ultraviolet (UV), while dynamically generating the Electroweak and Seesaw scales \textit{\'a l\'a} Coleman-Weinberg in the infrared (IR). This is particularly achieved by implementing threshold corrections at an energy scale $\mu_T$ arising due to the presence of vector-like fermions. We show that implementing the inflationary observables makes the couplings solely dependent on the plateau scale $M$, leaving us with only two independent parameters $M$ and $\mu_T$. Within the theoretically consistent parameter space defined by $m_{Z_{BL}} > 850~\rm GeV$, 
from the assumption of independent evolution of the dark sector couplings from the SM couplings and $M < 5.67~M_P$ required for the realisation of inflationary \textit{plateau-like} behaviour of the potential around $\phi=M$, where $M_P=2.4\times10^{18}$ GeV is the reduced Planck mass, we identify the parameter space that is excluded by the current LHC results from the search for the heavy $Z_{BL}$ boson. For typical benchmark points in the viable parameter regions, we estimate the reheating temperature to be $\mathcal{O}(TeV)$ thus consistent with the standard Big Bang Nucleosynthesis (BBN) constraints. For typical benchmark points ($M=5.67,~1,~0.1~M_P$) we predict the scales of inflation to be $\mathcal{H}_{inf}=2.79\times10^{12}$ GeV, $1.53\times10^{10}$ GeV and $1.53\times10^7$ GeV, respectively.}

\end{abstract}

\maketitle

\tableofcontents

\section{Introduction}

Grand Unified Theories (GUTs) formed the basis of the original proposal for the cosmic inflation, an accelerated expansion at the beginning of the universe, which can solve the horizon and the flatness problems as well as provide initial seed of density fluctuations to grow into our inhomogeneous universe as we see today \cite{PhysRevD.23.347,Sato:1980yn, 1980ApJ...241L..59K}, and later on, inflation was studied in the context of gravity effective theories like the Starobinsky scenario \cite{10.1143/PTP.46.776,STAROBINSKY198099}. Although the former turned out to be unsuccessful, the quantum generation of the primordial fluctuations seeding the large scale structure (LSS) of the Universe was a successful scenario. Irrespective of the origin of inflationary cosmology being of particle physics or not, the quite rapidly increasing data from cosmological precision measurements, particle physics experiments and astrophysical observations lead us to the quest of a coherent picture of the early Universe based on particle physics to begin with.

Although inflation can be achieved by scalar fields, particularly, slow-roll inflation by a single scalar field ($\phi$), simple potentials like quadratic $m^2\phi^2$ or quartic $\lambda_\phi \phi^4$ inflation scenarios which predict too large tensor-to-scalar ratio, have been ruled out by the observations of CMB power spectrum \cite{Akrami:2018odb}. Possibilities to rescue such models and make them consistent with the observations have been studied extensively in the literature. Whereas a possibility is to introduce non-minimal coupling of the inflaton to gravity ($\xi \phi^2 \mathcal{R}$) \cite{Bezrukov:2007ep, Libanov:1998wg, Fakir:1990eg, Futamase:1987ua, Masina:2018ejw,Okada:2010jf, Linde:1981mu, Kallosh:1995hi, Inagaki:2014wva} to flatten $\lambda_\phi \phi^4$ potential during inflation, the same purpose can also be achieved through Renormalization Group (RG)-improved $\lambda_\phi \phi^4$ potential. The quantum corrections generate a plateau shaped potential, whose flatness near the plateau (inflection point) makes the CMB constraints being satisfied. Particularly, employing bosonic and fermionic quantum corrections to achieve inflection-point inflation were studied in Refs. \cite{Okada:2017cvy, Choi:2016eif,Allahverdi:2006iq,Allahverdi:2006we,BuenoSanchez:2006rze,Baumann:2007np,Baumann:2007ah,Badziak:2008gv,Enqvist:2010vd,Cerezo:2012ub,Choudhury:2013jya,Choudhury:2014kma,Ballesteros:2015noa,Choi:2016eif,Okada:2016ssd,Okada:2017cvy,Stewart:1996ey,Stewart:1997wg,2104.03977}. In this set-up giving precise predictions in the CMB as well as long-lived particle searches \cite{Okada:2016ssd,Okada:2019yne,Okada:2019opp,1510.05669,1601.05979, 2008.09639,1706.09735,1902.02666, 0806.2798,1312.5944,1910.14586, 1910.14586} have been studied. In this paper we will investigate such a particle physics motivated scenario for inflation.

Any fundamental scalar field in quantum field theory (QFT) suffers from what is known as the hierarchy problem\footnote{Recently in higher-derivative non-local QFT scenarios, this problem can be relaxed, and conformal invariance can be dynamically achieved without introducing any new particles in the physical mass spectrum, see Refs. \cite{Ghoshal:2017egr,Ghoshal:2018gpq,Ghoshal:2020lfd,Frasca:2020jbe,Frasca:2020ojd,Frasca:2022duz} with predictions and interesting signals in LHC phenomenology \cite{Biswas:2014yia,Su:2021qvm}.}, although the Large Hadron Collider (LHC) was able to shed light on the origin of Electroweak Symmetry Breaking (EWSB) confirming the existence of a Brout-Englert-Higgs scalar doublet (commonly known as the Higgs doublet). In the SM, a non-zero vacuum expectation value (vev) of the Higgs doublet originates from a negative mass squared term in the Higgs potential at the tree level, which is the only mass term allowed by the symmetries of the SM. Quantum corrections to the mass of the SM Higgs doublet turn out to be UV sensitive, so that the effective Higgs mass is naturally be of the order of the Planck scale or the cut-off scale for the theory.

An elegant solution to this problem is to assume \textit{scale invariance}\footnote{We will use ``scale-invariance" and ``conformal invariance" inter-changeably in this paper, as they are known to be classically equivalent for any four-dimensional unitary and renormalizable field
theory perspectives \cite{Gross:1970tb,CALLAN197042,Coleman:1970je}.
}, and that all scales we observe be generated dynamically. One such attractive possibility was proposed long ago by Coleman and Weinberg, where a gauge symmetry breaking is dynamically (radiatively) generated via quantum corrections. However appealing as it may be, this mechanism fails within the Standard Model to generate the Higgs mass (the Electroweak Scale) because of the contributions of W and Z boson loops and from top quark loops. The original Coleman-Weinberg prediction was that the mass of the gauge bosons is greater than that of the Higgs boson,
$m_{Z,W} > m_H$ \cite{PhysRevD.7.1888,Englert:2013gz}.
In BSM scenarios, a picture that no scale is fundamental in nature and all mass scales are generated dynamically, has been explored extensively in the literature \cite{Adler:1982ri,Coleman:1973jx,Salvio:2014soa,Einhorn:2014gfa,Einhorn:2016mws,Einhorn:2015lzy}. In context of non-minimally coupling to gravity, such scenarios provide naturally flat inflaton potentials \cite{Khoze:2013uia,Kannike:2014mia,Rinaldi:2014gha,Salvio:2014soa,Kannike:2015apa,Kannike:2015fom,Barrie:2016rnv,Tambalo:2016eqr} and dark matter candidates \cite{Hambye:2013sna,Karam:2015jta,Kannike:2015apa,Kannike:2016bny,Karam:2016rsz,Barman:2021lot}, 
and also leads to very strong first-order phase transitions via supercooling in early universe and therefore the possibility of high amplitude detectable gravitational wave (GW) signals mainly due to dominance of thermal corrections in absence of tree-level mass terms~\cite{Jaeckel:2016jlh, Marzola:2017jzl, Iso:2017uuu, Baldes:2018emh, Prokopec:2018tnq, Brdar:2018num, Marzo:2018nov, Ghoshal:2020vud}. 
Scale invariant scenarios have always been seen as direction of model-building for the hierarchy problem in the Standard Model of particle physics
\cite{Foot:2007iy,AlexanderNunneley:2010nw,Englert:2013gz,Hambye:2013sna,Farzinnia:2013pga,Altmannshofer:2014vra,Holthausen:2013ota,Salvio:2014soa,Einhorn:2014gfa,Kannike:2015apa,Farzinnia:2015fka,Kannike:2016bny}. 
See Refs. \cite{1306.2329,1410.1817,0902.4050,0909.0128,1210.2848,1703.10924,1807.11490} for other studies of conformal invariance and dimensional transmutation of energy scales \cite{2012.11608,1812.01441,2201.12267,2011.10586,2010.15919,2102.10665,1812.02314,1709.09222}.

Sticking to the scale-invariant BSM framework, we consider a $B-L$ extended SM, in which the $B-L$ symmetry is broken by the Coleman-Weinberg mechanism, subsequently triggering the EW symmetry breaking. The $B-L$ extended SM \cite{Mohapatra:1980qe,Marshak:1979fm,Wetterich:1981bx,Masiero:1982fi,Mohapatra:1982xz} have been well-studied which accounts for matter-antimatter asymmetry and the origin of the SM neutrino masses via Type-I seesaw mechanism. Now as we go from UV to IR, quantum corrections from the $B-L$ gauge boson drives the running quartic coupling ($\lambda_\phi$) of the $B-L$ Higgs ($\phi$) negative in the IR. What happens is that, once the condition $\lambda_\phi<0$ is reached, $\phi$ develops a VEV $\phi=v_{BL}/\sqrt{2}$, and the mixing quartic term $-\lambda_{H\phi}|H|^2|\phi|^2$ with the SM Higgs doublet ($H$) effectively generates a negative mass squared, $-m^2|H|^2$  with $m^2=\frac{1}{2}\lambda_{H\phi}v_{BL}^2$, and hence the EW symmetry is broken.
Moreover, we consider a possibility that the $B-L$ Higgs is also responsible for the plateau inflation.

However, using the $B-L$ Higgs with conformal invariance as the inflaton for the radiative plateau inflation is highly non-trivial. Let us see in details why. In order for $B-L$ Higgs to drive the successful plateau inflation, the running of $\lambda_\phi(\phi)$ should exhibit a minima $\lambda_\phi(\phi=M)\sim0$, which means $\beta_{\lambda_\phi}<0$, $\beta_{\lambda_\phi}=0$ and $\beta_{\lambda_\phi}>0$ for $\phi<M$, $\phi=M$ and $\phi>M$, respectively, around $\phi=M$.
On the other hand, the CW mechanism requires $\lambda_\phi$ to be positive at high energies and fall to negative values at a low energy ($\phi\sim v_{BL}/\sqrt{2}$). Therefore, in order for the $B-L$ Higgs to play the dual role of the inflaton for the radiative plateau inflation and for breaking the $B-L$ symmetry by CW mechanism, $\beta_{\lambda_\phi}$ changes its sign twice from the UV to IR. It is non-trivial to realize such a behaviour of $\beta_{\lambda_\phi}$ in gauge theories.

In this paper, we propose a way to realize the dual role of the $B-L$ Higgs field, where threshold corrections from Majorana fermions at some intermediate scale play a crucial role \footnote{The mass of the fermions explicitly breaks the scale-invariance, but in this paper we do not go into details on the origin of the mass scale. However, we introduce such explicitly breaking terms only in the fermionic sector, so that no new hierarchy problem is created. See Refs. \cite{Brivio:2017dfq, Brdar:2018vjq} for such theories achieving electroweak symmetry breaking via radiative corrections.}. 

Our basic idea the following. At UV, to realize the successful plateau inflation, 
$\lambda_\phi(\phi=M)\sim0$ and its beta function is symbolically expressed as $\beta_{\lambda_\phi}\sim C_g g_{BL}^4-C_y y^4\sim 0$, where $C_{g,y}$ are numerical factors, and $y$ is the Yukawa coupling of a fermion with a mass $m_F$. Setting $g_{BL}$ and $y$ suitably, we can realize
$\beta_{\lambda_\phi}(\phi<M)<0$, $\beta_{\lambda_\phi}(\phi=M)=0$ and $\beta_{\lambda_\phi}(\phi>M)>0$, for generating the radiative plateau potential at $\phi=M$.
Moving towards the IR, the fermion decouples at $\phi=m_F$, and therefore, for $\phi<m_F$, the beta function changes its sign to $\beta_{\lambda_\phi}>0$. Moving to low energies further, $\lambda_\phi(\phi)$ eventually becomes negative and the $B-L$ symmetry by is broken by the CW mechanism.





The paper is arranged as follows: in section. \ref{inflation} we discuss inflation analysis of this work. In subsections \ref{basicinf},  \ref{model} and \ref{modelinf}, we discuss on the basics of plateau inflation, the model and on obtaining plateau inflation with the model. In section. \ref{cw all} we move to the Coleman-Weinberg part of this work. In subsections
\ref{CW},
\ref{diff inf+cw},
\ref{inf+cw} and 
\ref{theo cons},
we discuss the basics of Coleman-Weinberg mechanism, difficulty of achieving plateau inflation and Coleman-Weinberg in the same model, way out of this difficulty and theoretical conditions for this analysis to work respectively. We discuss the reheating analysis of this model in section. 
\ref{reheating}.
We discuss the possible searches for ${Z_{B-L}}$ particle and present the parameter space compatible with current experimental results in section.\ref{signatures}. In Section.\ref{conclusion}, we discuss the main findings of our paper and its implications. 

\medskip

\section{Inflation}\label{inflation}

In this section we will describe how to generate the inflationary plateau-like behaviour from $U(1)_{B-L}$ Higgs quartic potential, expanded around an inflection-point, driven purely due to radiative corrections.

\subsection{Basics of Inflection-point Inflation }
\label{basicinf}

Quickly re-capping the slow-roll parameters of inflationary observables,
\begin{eqnarray}
\epsilon(\phi)=\frac{ M_{P}^2}{2} \left(\frac{V'}{V}\right)^2, \; \; 
\eta(\phi)=
M_{P}^2\left(\frac{V''}{V }\right), \;\;
\varsigma^2{(\phi)} = M_{P}^4  \frac{V^{\prime}V^{\prime\prime\prime}}{V^2}, \label{SRCond}
\end{eqnarray}
where we have the reduced Planck mass $M_{P}= M_{Pl}/\sqrt{8 \pi} = 2.43\times 10^{18}$ GeV, $V$ is the inflaton potential, and the prime denotes its derivative with respect to the inflaton $\phi$.  

In this notation, the curvature perturbation $P ^2_{\mathcal{R}}$ is given by 
\begin{equation} 
P_{\mathcal{R}}^2 = \frac{1}{24 \pi^2}\frac{1}{M_P^4}\left. \frac{V}{ \epsilon } \right|_{k_0},
 \label{PSpec}
\end{equation}
 the value of which should be $P_\mathcal{R}^2= 2.189 \times10^{-9}$
  from the Planck 2018 results \cite{Akrami:2018odb} at pivot scale $k_0 = 0.05$ Mpc$^{-1}$. 
The number of e-folds is given by,
\begin{eqnarray}
N=\frac{1}{M_{P}^2}\int_{\phi_E}^{\phi_I}\frac{V }{V^\prime} d\phi  ,
 \label{EFold}
\end{eqnarray} 
where $\phi_I$ is the value of inflaton during horizon exit of the scale $k_0$, 
  and $\phi_E$ is the value of inflaton value when the slow-roll condition is violated, 
  i.e. $\epsilon(\phi_E)=1$. The slow-roll approximation holds whenever
   $\epsilon \ll 1$, $|\eta| \ll 1$ , and $\varsigma^2\ll1$.

The inflationary predictions of the scalar and tensor perturbations are given by,
\be
n_s = 1-6\epsilon+2\eta, \; \; 
r = 16 \epsilon , \;\;
\alpha = 16 \epsilon \eta -24 \epsilon^2-2 \varsigma^2, 
 \label{IPred}
\ee
where $n_{s}$ and $r$ and $\alpha \equiv \frac{\mathrm{d}n_s}{d ln k}$ are the scalar spectral index, the tensor-to-scalar ratio and the running of the spectral index, respectively, at $\phi = \phi_I$.  
Planck 2018 results give an upper bound on  $r \lesssim 0.067$,  
bound for the spectral index ($n_s$) and the running of spectral index ($\alpha$) to be $0.9691 \pm 0.0041$ and $0.0023\pm 0.0063$, respectively \cite{Akrami:2018odb}. A combination of Planck with BICEP/Keck 2018 and Baryon Acoustic Oscillations data tightens the upper bound of tensor-to-scalar ratio to $r < 0.032$ \cite{Tristram:2021tvh}.\footnote{As we will see later, the highest $r$ value that maybe achievable in this model is for a benchmark point $M=5.67M_P$, which corresponds to $r=0.00012$.}


The inflaton potential for inflection-point inflation, expanding around an inflection point near $\phi = M$ value of the field is given by \cite{Okada:2016ssd}:
\be
V (\phi)\simeq V_0 +\sum_{n = 1}^3 \frac{1}{n!}V_n (\phi-M)^n , 
\label{eq:PExp}
\ee
   where $V_0 = V(M)$ is constant, $V_n \equiv  d^{n}V/ d \phi^n |_{\phi =M}$ are derivatives evaluated at at $\phi=M$, and the inflection-point $\phi = M$ is the field value at the pivot scale $k_0= 0.05$ Mpc$^{-1}$ of the Planck 2018 measurements \cite{Akrami:2018odb}. If the values of $V_1$ and $V_2$
   are tiny enough, inflection-point can be realized. re-writing Eqs.~(\ref{SRCond}) in terms of parameters of Eq. (\ref{eq:PExp}), 
\bea
\epsilon(M) \simeq \frac{M_{P}^2}{2}\Big(\frac{V_1}{V_0}\Big)^2, \;\;
\eta(M) \simeq M_{P}^2\Big(\frac{V_2}{V_0}\Big), \;\;
\zeta^2{(M)} = M_{P}^4  \frac{V_1 V_3}{V_0^2}, 
\label{IPa}
\eea
where we have used the approximation $V(M)\simeq V_0$. 
Similarly, the power-spectrum $\Delta_{\mathcal{R}}^2$ is expressed as
\be
P_{\mathcal{R}}^2 \simeq \frac{1}{12\pi^2}\frac{1}{M_P^6}\frac{V_0^3}{V_1^2}.
\label{CV1} 
\ee
Using the observational constraint, $P_{\mathcal{R}}^2= 2.189 \times  10^{-9}$, and a fixed $n_s$ value,
 we obtain 
\bea
\frac{V_1}{M^3}&\simeq& 1963\left(\frac{M}{M_P}\right)^3\left(\frac{V_0}{M^4}\right)^{3/2}, \nonumber \\
\frac{V_2}{M^2}&\simeq& -1.545\times 10^{-2}\Big(\frac{1-n_s}{1-0.9691}\Big)\Big(\frac{M}{M_P}\Big)^2\left(\frac{V_0}{M^4}\right), 
\label{FEq-V1V2}
\eea
using $V(M)\simeq V_0$. 
For the remainder of the analysis we set $n_{s}=0.9691$ (the central value from the Planck 2018 results \cite{Akrami:2018odb}). 
Then V$_3$ becomes
\bea
\frac{V_3}{M} \simeq 6.983 \times 10^{-7}\Big(\frac{60}{N}\Big)^2 \Big(\frac{V_0^{1/2}}{M M_P}\Big) . 
\label{FEq-V3} 
\eea

\smallskip

Using Eqs.~(\ref{IPred}), (\ref{IPa}), (\ref{FEq-V1V2}) and (\ref{FEq-V3}), 
 the tensor-to-scalar ratio ($r$) is given by
\bea 
r=3.082\times 10^7\Big( \frac{V_0}{M_P^4}\Big). 
\label{FEq-r}
\eea 
and, the running of the spectral index ($\alpha$) 
\bea
\alpha \simeq - 2\varsigma^2(M) = - \; 2.741 \times 10^{-3}\left(\frac{60}{N}\right)^2, 
\label{FEq-alpha}
\eea
It is interesting to note that the running is independent of the $V_0$ and $M$ terms in the inflation potential.

This prediction is consistent with the  current experimental bound, $\alpha=0.0023\pm 0.0063$ \cite{Akrami:2018odb}. 
Precision measurement of the running of the spectral index in future experiments can reduce the error to $\pm0.002$ \cite{Ade:2018sbj,Aiola:2020azj}. 
Hence, the predictions can be tested in the future.


\subsection{The $B-L$ Extended Model}\label{model}



{
\renewcommand{\arraystretch}{1.4}
\begin{table}[H]
\centering
\begin{tabular}{ccc} 
\hline \hline 
Field & Group & Coupling \\ 
 \hline 
$Z_{BL}~~~~$ & $U(1)_{B-L}$ & $g_{BL}$ \\
\hline \hline
\end{tabular} 
\caption{New gauge sector of the model 
}
\label{tab:DarkBS1}
\end{table}

\begin{table}[H]
\centering
\begin{tabular}{ccccc} 
\hline \hline 
Field & Spin  & \( U(1)_{B-L} \) \\ 
\hline
\(~~~~\phi~~~~\) & \(~~~~0~~~~\)  & \(2 \) \\ 
\(~~~~\psi_{L,R}~~~~\) & \(~~~~\frac{1}{2}~~~~\)  & \(-1 \) \\ 
\(~~~~N_R^i~~~~\) & \(~~~~\frac{1}{2}~~~~\)  & \(-1 \) \\ 
\hline \hline
\end{tabular} 
\caption{New scalars and fermions in the model.}
\label{tab:DarkBS2}
\end{table}

}


We start with the minimal extension of the SM model where the SM gauge group is supplemented with a $U(1)_{B-L}$ local symmetry. In this $B-L$ sector, there are two vector-like fermions $\psi_L$ and $\psi_R$, three right-handed neutrinos $N_R^i$, as well as the complex scalar field $\phi$. Tables
\ref{tab:DarkBS1} and \ref{tab:DarkBS2} show all the details about the
gauge sector and the new scalars and fermions in the model.
%
%
 The vector-like fermions $\psi_L$, $\psi_R$ and $N_R^i$ are  charged under $U(1)_{B-L}$. In addition to the usual canonical kinetic energy terms, the $B-L$ Higgs interacts with right-handed neutrinos ($N_R^i$) and vector-like fermions ($\psi_L$ and $\psi_R$) through Yukawa interaction terms, 
\bea  
   {\cal L} \supset  - \frac{1}{2} \sum_{i=1}^{3} Y_i^{low}  \phi  \overline{N_{R}^{i~C}} N^i_{R} - \frac{1}{2}  y_L  \phi  \overline{\psi_L^{C}} \psi_L
    - \frac{1}{2}  y_R  \phi  \overline{\psi_R^{C}} \psi_R+{\rm h.c.},
   \label{eq:potyukawa}
\eea

and with SM Higgs through,
\begin{equation}
\mathcal V\left( H , \phi \right) =  \lambda_{H} |H| ^4 - \lambda_{H \phi} \, |H|^2 |\phi|^2 + \lambda_{\phi} |\phi| ^4.
\label{eq:potscalar}
\end{equation}
The choice of negative sign before the mixing term $\lambda_{H \phi}  |H|^2 |\phi|^2$ will be explained in section. \ref{CW}. For the analysis in this work we assume that the contribution of $\lambda_{H\phi}$ to be negligible with respect to that of $g_{BL}$ in $\beta_{\lambda_\phi}\equiv \phi \frac{d \lambda_\phi}{d \phi}$, i.e. $8\lambda_{H\phi}^2<96g_{BL}^4$. This simplifying assumption lets us study the running of the couplings in the dark sector and that of the SM sector independently of each other. This condition will be discussed in more details in section. \ref{theo cons}.
 For this assumption, the couplings of the dark sector follow the RG equations,
\bea
16 \pi^2 \phi  \frac{d g_{BL}}{d \phi} &=& \left(12+\frac{4}{3}\right) g_{BL}^3,         \nonumber\\
 16 \pi^2\phi \frac{d Y_i^{low}}{d \phi}   &=& 6 g_{BL}^2 Y_i^{low}+Y_i^{low}\left(\frac{1}{2}\left( \sum_j Y_j^{low~2}+y_L^2+y_R^2\right)-12 g_{BL}^2+Y_i^{low~2}\right),   \nonumber\\
 16 \pi^2\phi \frac{d y_L}{d \phi}   &=& 6 g_{BL}^2 y_L+y_L\left(\frac{1}{2}\left( \sum_j Y_j^{low~2}+y_L^2+y_R^2\right)-12 g_{BL}^2+y_L^2\right),   \nonumber\\
 16 \pi^2\phi \frac{d y_R}{d \phi}   &=& 6 g_{BL}^2 y_R+y_R\left(\frac{1}{2}\left( \sum_j Y_j^{low~2}+y_L^2+y_R^2\right)-12 g_{BL}^2+y_R^2\right),   \nonumber\\
 16 \pi^2\phi \frac{d \lambda_\phi}{d \phi}  &=& 20 \lambda_\phi^2+96 g_{BL}^4-\left( \sum_j Y_j^{low~4}+y_L^4+y_R^4\right)\nonumber\\
 &{}&+ \lambda_\phi\left( 2\sum_j Y_j^{low~2}+2y_L^2+2y_R^2- 48 g_{BL}^2\right).
  \label{RGEs}
\eea
For simplicity, we consider degenerate Yukawa couplings, $ y_L = y_R \equiv y$ and $ Y_1^{low}=Y_2^{low}=Y_3^{low} \equiv Y^{low}$ throughout this work. We also choose $Y_i^{low}\ll y_{L,R}$ and $Y_i^{low}\ll g_{BL}$ for simplicity and ignore the $Y_i^{low}$ terms in the next section. It is worth mentioning that, although the right handed neutrinos $N_R^i$ and the couplings $Y_i^{low}$ do not affect the analysis in this work due to their assumed smallness, after the breaking of the $B-L$ symmetry they naturally obtain mass $m_{N_R^i} \sim Y_i^{low} v_{BL}$, $v_{BL}$ being the $B-L$ vev. Hence the Seesaw scale is dynamically generated.

%
%
%


\subsection{Achieving Inflection-point in the UV in the Model}
\label{modelinf}

For achieving the radiative plateau in the U(1)$_{B-L}$ Higgs potential $V_{tree}= (1/4)  \lambda_{\phi-tree} \phi^4$, we go to the RGE-improved effective potential,  
\bea
V(\phi) = \frac{1}{4} \lambda_\phi (\phi)\;\phi^4, 
\label{VEff}
\eea
where $\lambda_\phi (\phi)$ is the solution to the RGE, as in Eqs. \ref{RGEs}, which involves $g_{BL}$, $y$ and $\lambda_\phi$. 
The coefficients in the expansion of Eq.~(\ref{eq:PExp}) in term of the model parameters is given as \footnote{ See Refs. \cite{Okada:2016ssd} for detailed derivation.}, 
\bea
\frac{V_1}{M^3}&=& \frac{1}{4} (4 \lambda_\phi + \beta_{\lambda_\phi}),\nonumber \\
\frac{V_2}{M^2}&=& \frac{1}{4} (12\lambda_\phi + 7\beta_{\lambda_\phi}+M \beta_{\lambda_\phi}^\prime), \nonumber \\
\frac{V_3}{M}&=& \frac{1}{4} (24\lambda_\phi + 26\beta_{\lambda_\phi}+10M \beta_{\lambda_\phi}^\prime+M^2 \beta_{\lambda_\phi}^{\prime\prime}), 
\label{ICons2}
\eea
where the prime denotes differentiation with respect to the field $\phi$, i.e., $d/d\phi$.
For the condition of inflection-point, using $V_1/M^3\simeq 0$ and $V_2/M^2\simeq 0$, we obtain $V_3/M \simeq 16 \;\lambda_\phi(M)$, which in turn when compared to Eq.~(\ref{FEq-V3}), gives,
\bea
\lambda_\phi(M)\simeq 4.762 \times 10^{-16} \Big(\frac{M}{M_{P}}\Big)^2\Big(\frac{60}{N}\Big)^4,
\label{FEq1} 
\eea 
where we have approximated $V_0\simeq (1/4) \lambda_\phi(M) M^4$. 
Since $\lambda_\phi(M)$ is extremely small, we can approximate $\beta_{\lambda_\phi}(M) \simeq 0$ at one-loop level\footnote{In perturbation theory, two loop contributions to the beta functions are is subdominant than one loop contributions. So, once we make sure that the one-loop corrections corresponding to the most dominant contributions cancel out, the deviation from this cancellation becomes less and less severe when we take into account the higher-loop contributions. So, our results obtained by the requirement of cancellation at one-loop level are not significantly altered by the higher order corrections.} , leading to, 
\be
y(M)\simeq \left(\frac{96}{2}\right)^{1/4}\;g_{BL}(M), 
\label{FEq3}
\ee
This equation implies that, to realize a successful inflection-point inflation, we need a fixed ratio between the mass of the vector like fermions and the gauge boson mass\footnote{We again emphasize that, to realise a successful inflection point inflation, the tuning of the parameters is necessary. This is a general problem for inflection point inflation scenarios. However we want to point out the interesting property that, this cancellation leads to a relation between otherwise unrelated couplings.}. 
Using $V_2/M^2\simeq 0$ 
and Eq.~(\ref{FEq3}), we find $\lambda_\phi(M)\simeq 5.27\times 10^{-3} \;g_{BL}(M)^6$. 
Then using Eq.~(\ref{FEq1}), $g_{BL}(M)$ can be written as
\be
g_{BL}(M)\simeq 6.701\times 10^{-3} \;\Big(\frac{M}{M_{P}}\Big)^{1/3}.
\label{FEq2} 
\ee
Finally, from Eqs.~(\ref{FEq-r}) and (\ref{FEq1}), the tensor-to-scalar ratio ($r$) is given by 
\be
r \simeq 3.670 \times 10^{-9}  \Big(\frac{M}{M_{P}}\Big)^6, 
\label{FEqR} 
\ee
which is extremely small, as expected for the single field inflationary scenario where the potential is flat at the pivot scale. 


It is important to mention that the theoretical consistency of this analysis depends on the fact that $V_3$ is dominant over any $V_n$ for all other $n$ values in Eq. \ref{eq:PExp} to realize the plateau-like behaviour. This condition, to be precise, $V_3 > V_4$, leads to the upper limit \cite{Okada:2016ssd}:
\be
\label{Mupper_bound}
M < 5.67~M_P
\ee

\begin{figure}[H]
\begin{center}
\includegraphics[height=0.49\textwidth]{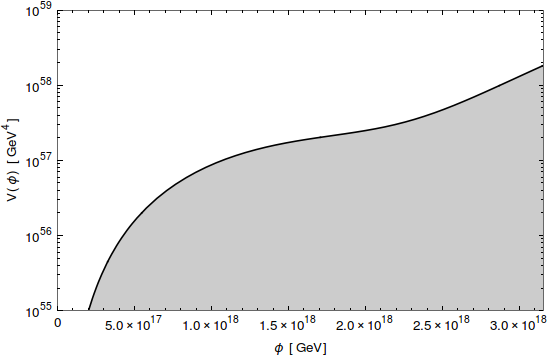}
	\caption{\it RGE-improved inflaton quartic potential plotted against $\phi$. 
	$M = M_P = 2.43 \times 10 ^{18}$ GeV.
	 }
\label{fig:potinf}
\end{center}
\end{figure}

Just to show an example, for the choice of $M=M_P$, we get from the analysis done earlier in this section, $g_{BL}= 6.701\times10^{-3},~y=0.0176,~\lambda_\phi=4.76\times10^{-16}$ during inflation, 
the potential for this choice of parameters is shown in Fig. \ref{fig:potinf}.




We note a crucial point at the end of this section to carry forward to the next section: that the inflection-point and the inflationary plateau-like behaviour necessarily leads to $\lambda_{\phi} \simeq 0$ and $\beta_{\lambda_{\phi}} \simeq 0$ at M (UV scale) which means that in our model due to the presence of the Yukawa coupling $y$, $\lambda_{\phi}$ monotonically grows in the IR from UV. In the next section, we will see that for the requirement of radiative symmetry breaking via CW mechanism, we will also need positive contribution in $\beta_{\lambda_\phi}$ to dominate in order to have negative $\lambda_\phi$ in the IR.

\medskip

\section{Coleman-Weinberg in the IR}
\label{cw all}
In this section we present the Coleman-Weinberg pathway to generate the EW and Seesaw scales in the IR. Such a scenario demands as we will show to be difficult to achieve the inflationary-plateau behaviour as described in the last section. We will present a possible resolution to the problem by implementing energy threshold correction and derive the full sets of conditions on the parameters of the model that lead to the whole picture to be consistent.

\subsection{Dynamical Generation of EW and $B-L$ Seesaw Scales}
\label{CW}

As discussed in section. \ref{model}, we  work with gauge group $\rm U(1)_Y \times SU(2)_L \times SU(3)_c \times U(1)_{B-L}$, namely the SM gauge group with an extra $U(1)_{B-L}$. At low energy, we have dynamical symmetry breaking down to ${\rm U}(1)_{\rm em}\times SU(3)_c$ gauge group. However, as we do not have any dimensionful parameter in the Lagrangian, we depend on dynamical generation of scales through running of couplings. To discuss the basics of this dynamical generation, we consider $\beta_{\lambda_\phi}$ in Eq. \ref{RGEs}. If there is no fermionic contribution, we get,
\be
\phi \frac{d \lambda_\phi}{d \phi}  = \frac{1}{16 \pi^2} \left(20 \lambda_\phi^2+96 g_{BL}^4
-48 \lambda_\phi  g_{BL}^2\right).
\ee
As a consequence, when there is no negative Yukawa contribution, $\beta_{\lambda_{\phi}}$ is positive definite and the $g_{BL}$ term makes $\lambda_{\phi}$ negative at low energy, leading to dynamically generated vev of $\phi$, $ v_{BL}$.


The effective scalar potential at one loop order can be approximated by inserting a running
$\lambda_{\phi}$ in the tree-level potential of Eq.\ref{VEff}:
\be
V (\phi) \simeq \frac{1}{4} \lambda_{\phi} \phi^4 \simeq  \frac{1}{4} \beta_{\lambda_\phi} \ln (\phi/\phi_*) \phi^4  ,
\ee
where $\phi_*$ is the critical scale below which $\lambda_{\phi}$ becomes negative. The potential attains a minima as $\lambda_{\phi}$ is negative at small energy scales, as discussed latter in section, \ref{reheating}.  Once $\phi$ attains a vev, the mixing $-\lambda_{H \phi}  |H|^2 |\phi|^2$ works as the symmetry breaking term of SM Higgs, the mass matrix analysis of the system will be discussed latter in section. \ref{reheating}.

Effectively we can say that $\phi$ acts as `the Higgs of the $B-L$ Higgs' and as `the Higgs of Seesaw Scale'.
Furthermore, the `Higgs of the $\phi$' is $\phi$ itself, i.e.
the EW scale and the seesaw scale are dynamically generated via dimensional transmutation when  $U(1)_{B-L}$ symmetry is broken radiatively, i.e. $\lambda_\phi < 0$.

\subsection{Difficulty with Inflection-point Conditions}
\label{diff inf+cw}

However, as discussed in section. \ref{inflation}, at the scale of plateau inflation $\phi=M$, $\beta_{\lambda_\phi} \simeq 0$ due to cancellation of the bosonic and fermionic contributions, i.e. $\beta_{\lambda_\phi} \simeq 96 g_{BL}^4-2 y^4=0$,  
considering the dominant contributions in Eq. \ref{RGEs}. As the inflaton field $\phi$ rolls down the potential to scales below the scale of the inflationary plateau around M, the fermionic contributions $\sim y^4$ becomes dominant and $\beta_{\lambda_\phi}<0$. So, it seems impossible to make $\lambda_\phi$ negative in the IR again just using radiative corrections.

\subsection{Achieving the Inflationary Plateau via Threshold Correction}
\label{inf+cw}

One possible resolution to achieve the inflection-point at UV and $\lambda_{\phi} < 0$ at IR simultaneously in a model maybe possible through threshold energy corrections, i.e. going from UV to IR, if at some scale, say $\mu_T$, a fermionic contribution $2 y^4$ to $\beta_{\lambda_\phi}$ vanishes, the bosonic contribution $96 g_{BL}^4$ again dominates, making $\beta_{\lambda_\phi}>0$ again in the IR. This makes it possible for $\lambda_\phi <0 $ and forces the $U(1)_{B-L}$ symmetry breaking, \textit{\'a l\'a} CW mechanism.

\begin{figure}[H]
\begin{center}
\includegraphics[height=0.43\textwidth]{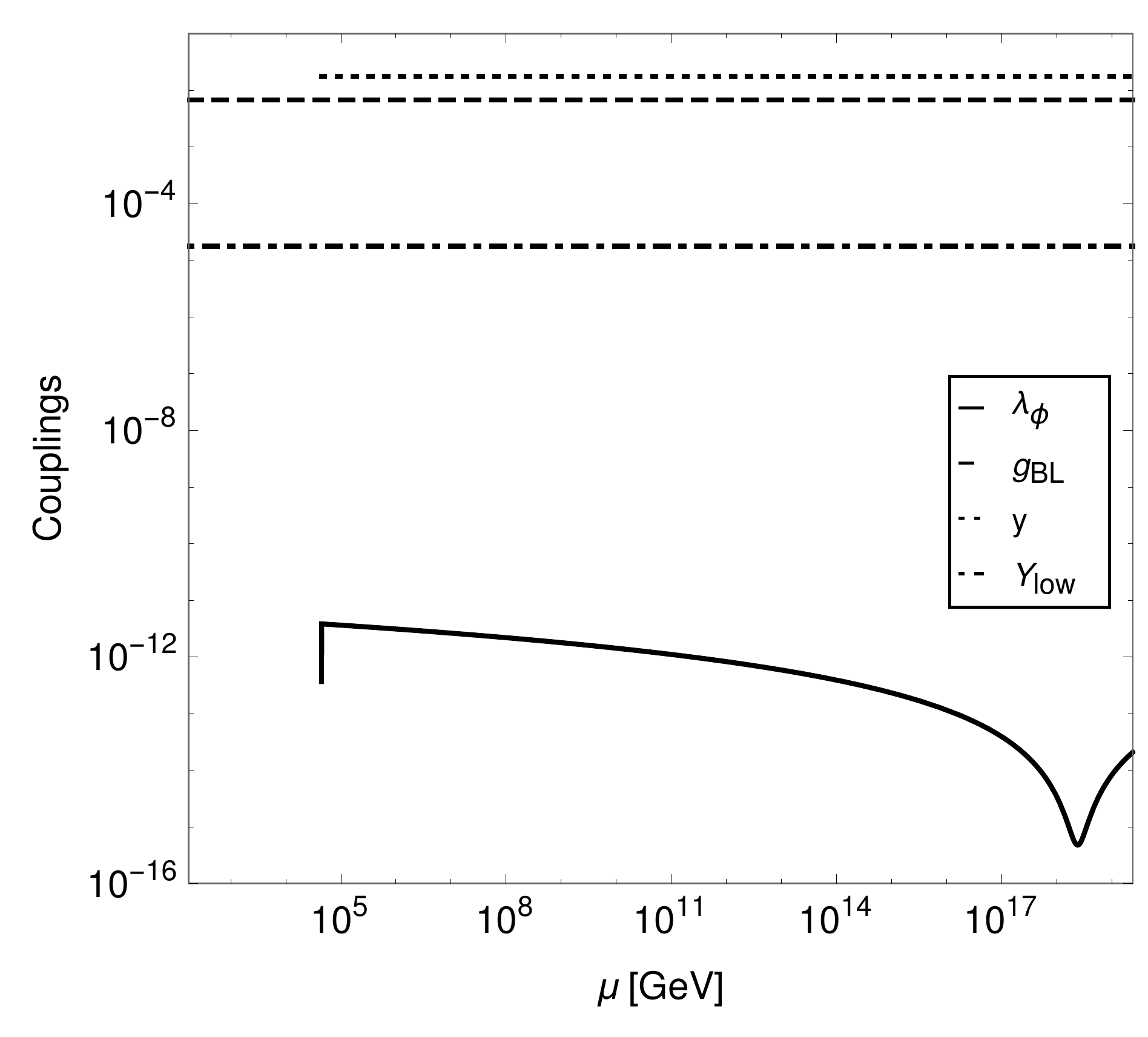}
\includegraphics[height=0.43\textwidth]{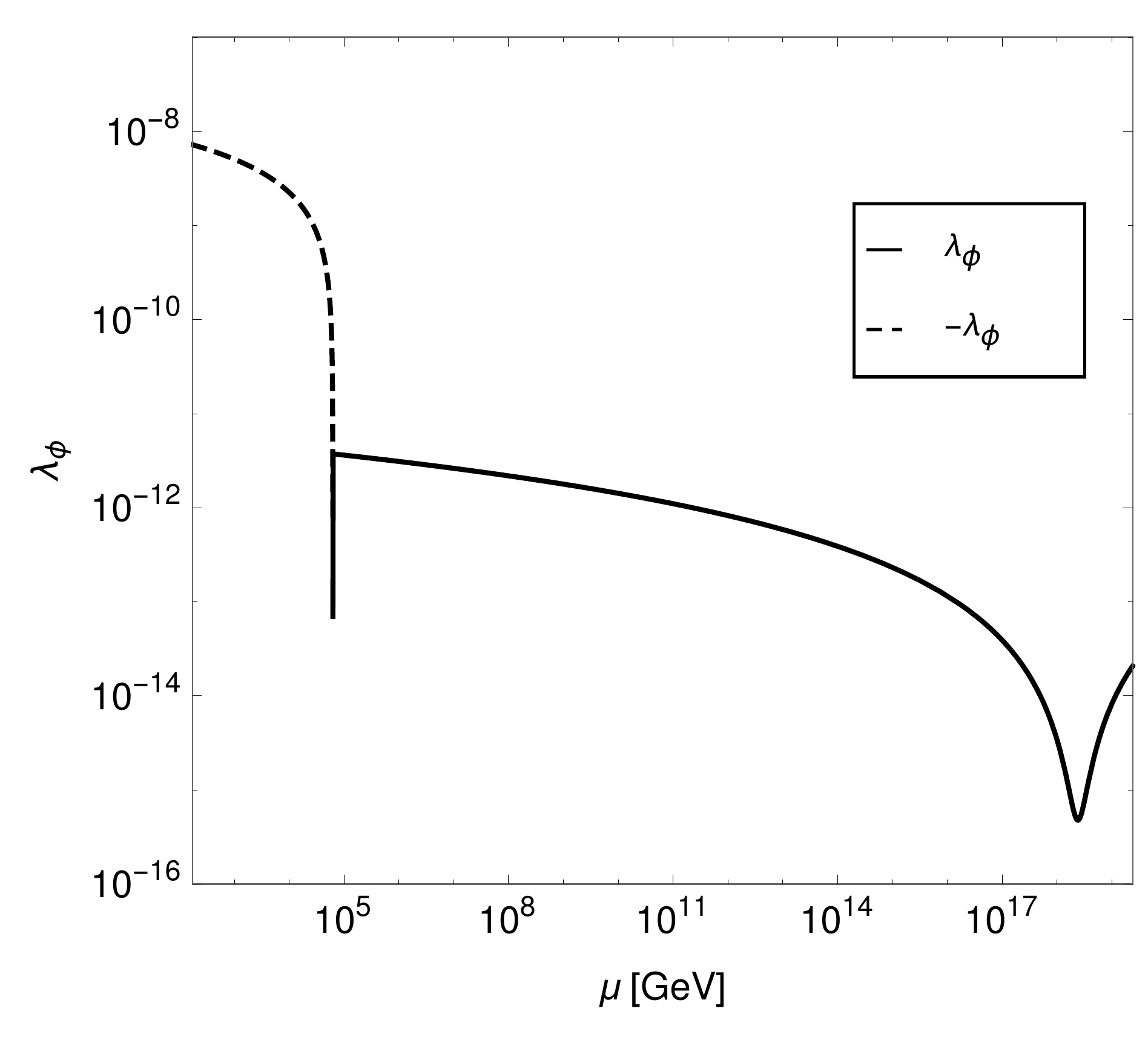}
	\caption{\it \textbf{Left Panel:} RG running of all the couplings for the benchmark point (M = 1 $M_P$, $\mu_T = 44.85$ TeV) against $\mu$. \textbf{Right Panel:} RG running of $\lambda_\phi$ against $\mu$. Note the abrupt drop of $\lambda_\phi$ to negative value at the threshold. We have chosen negligible $Y^{low}=10^{-3} y$ for this work.}
\label{fig:rgeall}
\end{center}
\end{figure}

We lay out the prescription for this mechanism in the model \ref{model}: in our model the additional $B-L$ Higgs $\phi$ as the inflaton interacts with right-handed neutrinos ($N_R^i$) and vector-like fermions ($\psi_{L,R}$) through Yukawa interaction terms, $Y_j^{low},~y_L$ and $y_R$ respectively, as per Eq. \ref{eq:potyukawa} in the UV but stop affecting the RGE-improved potential in the IR due to threshold correction, at a scale $\mu_T$; RGE below the threshold energy scale $\mu=\mu_{T}$ becomes (eliminating the contribution of $y_L$ and $y_R$ in Eq. \ref{RGEs}),
\bea
\phi \frac{d \lambda_\phi}{d \phi}  &=& \frac{1}{16 \pi^2} \left(20 \lambda_\phi^2+96 g_{BL}^4-\sum_j Y_j^{low~4}
 + \lambda_\phi\left( 2\sum_j Y_j^{low~2}- 48 g_{BL}^2\right)\right),
  \label{RGEs2}
\eea
Besides the scale of inflection-point M, we have the threshold scale $\mu_T$ as the only free parameter in our model. We shall see later that $\mu_T$ has a lower bound depending on $g_{BL}$, hence $M$, if we assume that RGEs of Dark sector and SM sector evolve independently. If $Y_i^{low} \ll y_{L,R}$ and $Y_i^{low}\ll g_{BL}$, the value of $\lambda_\phi$ suddenly drops to negative value below  $\mu_T$ ( as shown in Fig. \ref{fig:rgeall}).
The sudden drop is due to the fact that $-48\lambda_\phi g_{BL}^2$ is negligibly small with respect to $96g_{BL}^4$ (see Eq.\ref{RGEs2}). If the difference between the contributions $-48\lambda_\phi g_{BL}^2$ and $96g_{BL}^4$ is smaller, we may achieve a smooth transition of $\lambda_\phi$ from positive to negative values. For the choice of parameters $\mu_T=44.85$ TeV \footnote{This choice of $\mu_T$ corresponds to the  intersection of ATLAS final result constraint on $g_{BL}$ vs $m_{Z_{BL}}$ plane and the line corresponding to $M=M_P$, as discussed later and shown in Fig. \ref{fig:atlas}. In general we choose $\mu_T$ as the maximum of $\mu_T$ values corresponding to the intersection point described above and the intersection of lower bound of $m_{Z_{BL}}$ (for theoretical consistency) with the line of constant $M$.}, $M=M_P$ and choosing negligible $Y^{low}=10^{-3} y$, the evolution of the couplings are shown in Fig. \ref{fig:rgeall}.

In summary, to start with we had six parameters in our model, namely the scale of inflationary plateau $M$, the threshold scale $\mu_T$, and the couplings $\lambda_\phi,~g_{BL},~y,~Y^{low}$. Among them we choose $Y^{low}\ll y$ and $Y^{low}\ll g_{BL}$, such that they do not affect the RGE-improved inflationary potential V($\phi$) thus leaving us with 5 parameters. Fixing the observed values of the inflationary parameters $P ^2_{\mathcal{R}}$, $n_s$ and a chosen value of e-foldings $N$ makes the couplings $\lambda_\phi,~g_{BL},~y$ at $\phi=M$ dependent only on $M$, therefore reducing the system to only two independent parameters $M$ and $\mu_T$ in totality in our model.

\subsection{Conditions for Theoretical Consistencies}
\label{theo cons}

In order to simplify our model we assume that RGEs of the dark sector and SM sector evolve independently of each other, as previously mentioned in section. \ref{model}. To satisfy this assumption we require that the positive contribution of $\lambda_{H\phi}^2$ in $\beta_{\lambda_\phi}$ to be negligible w.r.t the contribution from gauge coupling $g_{BL}^4$ in Eq. \ref{RGEs}, i.e. $8\lambda_{H\phi}^2<96g_{BL}^4$. In the other extreme of this assumption, i.e. for $8\lambda_{H\phi}^2>96g_{BL}^4$, the inflection point is achieved via cancellation of contributions from $\lambda_{H\phi}^2$ and $y^4$ in $\beta_{\lambda_\phi}$, however we do not discuss this scenario in this work (see \cite{Caputo_2019} for such an example). Now, the term $-\lambda_{H \phi} \, |H|^2 \phi^2$ in the scalar potential, 
when 
$\phi =\langle\phi\rangle= v_{BL}/\sqrt{2}$, works as $-m^2 |H|^2$ (where $m=m_h/\sqrt{2}=88$ GeV, $m_h=125$ GeV being the mass of SM Higgs) to provide spontaneous symmetry breaking of SM Higgs, so we get $m_h^2 = \lambda_{H \phi} v_{BL}^2$.
As for our case $\mu_T \sim v_{BL}/\sqrt{2}$ (due to the sharp drop of $\lambda_{\phi}$ at $\mu_T$), we get a lower limit of $\mu_T$, i.e. 
\be
\mu_T\gtrsim \frac{v_{BL}}{\sqrt{2}}=\frac{m}{\sqrt{\lambda_{H \phi}}}>\frac{m}{(2 \sqrt{3} g_{BL}^2)^{1/2}}.
\ee

This condition also gives us a lower limit on $m_{Z_{BL}}$,
\be
m_{Z_{BL}}=2g_{BL}v_{BL}>\frac{2\sqrt{2}g_{BL}m}{(2 \sqrt{3} g_{BL}^2)^{1/2}}=\frac{2}{3^{1/4}}m=133 ~\rm GeV
\ee

In terms of \textit{order-of-magnitude} estimate, this will be truly valid for $8\lambda_{H\phi}^2 \ll 96g_{BL}^4$, setting $8\lambda_{H\phi}^2$ to be $\lesssim10 \time 96g_{BL}^4$, requiring $m_{Z_{BL}}>3\times133 ~\rm GeV$. This we chose to be our lower bound on $ m_{Z_{BL}}$. This condition translated to $M$ as the lower bound for some chosen value of $\mu_T$, along with the upper bound condition from Eq.\ref{Mupper_bound} will assure us the theoretical consistency of our analysis and results.

Along with this bound of $m_{Z_{BL}}>3\times133 ~\rm GeV$, we have a stronger bound on $m_{Z_{BL}}$ from the theoretical consistency of our analysis, as will be discussed in the section \ref{reheating}. This bound comes from the calculation of masses of the scalar mass eigenstates. The constraint we use in this work is,
\be
m_{Z_{BL}}\gtrsim850 ~\rm GeV.
\ee


\section{Reheating the Visible Universe}
\label{reheating}





\begin{figure}[H]
\begin{center}
\includegraphics[height=0.43\textwidth]{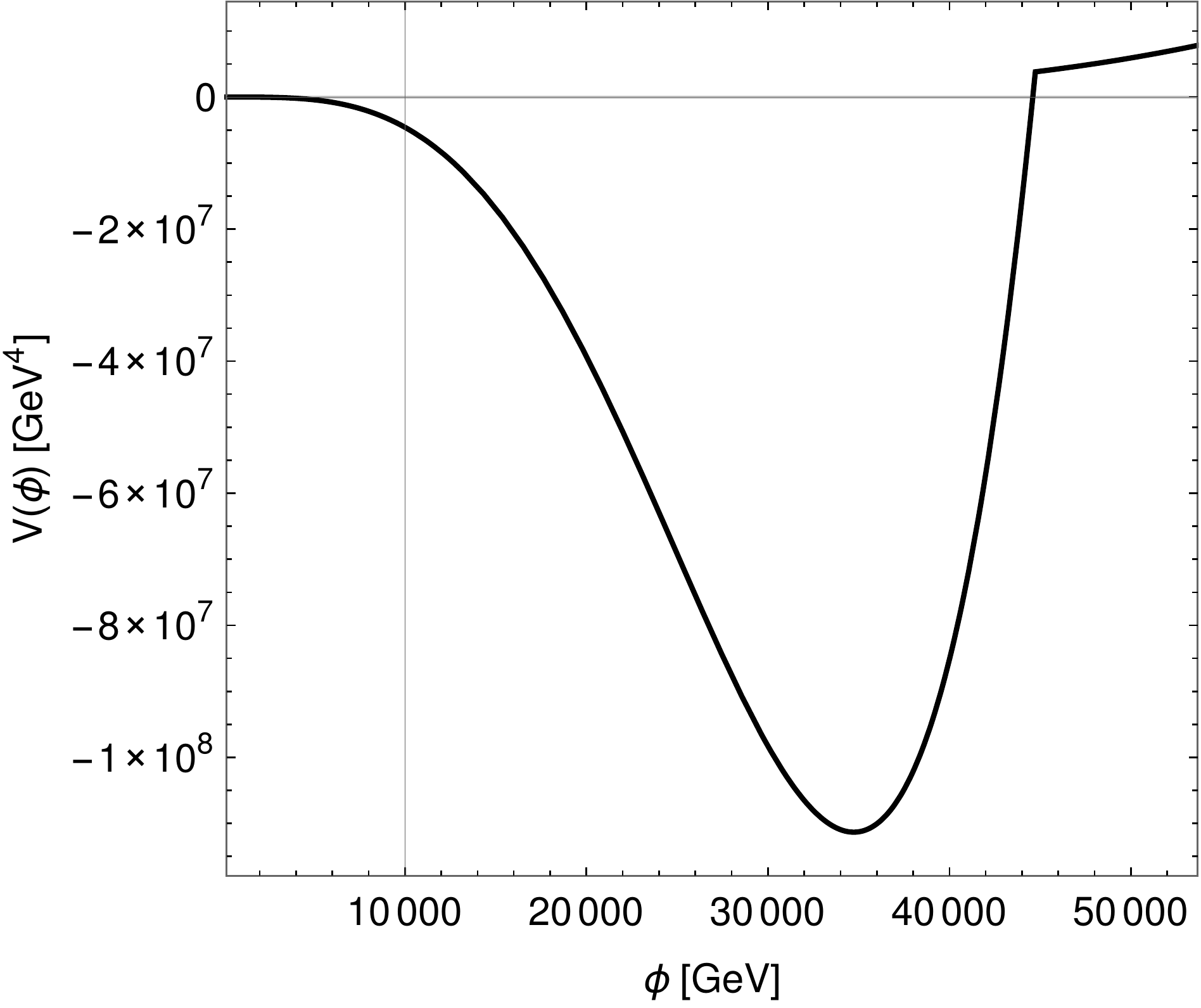}
	\caption{\it Plot of $V(\phi)$ for the benchmark point (M = 1 $M_P$, $\mu_T = 44.85$ TeV) near the threshold scale and minima of the potential in IR. Note that due to the sharp drop in potential to the minima near $\phi=\mu_T$, we can use the approximation $v_{BL}/\sqrt{2}\sim \mu_T$. }
\label{fig:minphi}
\end{center}
\end{figure}

Now let us turn towards the reheating dynamics to connect our model with the standard Big Bang Cosmology. This occurs when inflation has terminated and the inflaton oscillates around the minima of its potential, interpreted as a collection of particles at rest and decays perturbatively. The reheating temperature $T_{rh}$ is then given by\footnote{If the inflaton couples to other fields with sizable couplings, it may indeed give rise to significant energy transfer to those sectors via preheating. However, it is difficult for preheating to transfer the total energy from the inflaton to the radiation, and some energy density is left in the inflaton field. In our scenario, the inflaton potential around the minimum behaves as a quadratic potential, and the left-over energy density stored in the inflaton arising due to oscillation around the minimum, behaves like the equation for state for matter. Thus, the radiation energy density produced during preheating dilutes away with respect to the inflaton energy density, unless the inflaton decays right after preheating. So, we think it reasonable to estimate the reheating temperature by the perturbative decay rate of the inflaton. Moreover, in our case, the main reheating process is via fermion production from the inflaton, since the gauge boson $Z'$ is heavier than inflaton. We expect the parametric resonance to be further suppressed via Pauli Blocking.},
\be
T_{rh}\simeq .55\left(\frac{100}{g_*}\right)^{1/4}\sqrt{\Gamma_\varphi M_P};
\ee
where $\Gamma_\varphi$ is the decay rate of the redefined inflaton $\varphi=\sqrt{2}\texttt{Re}(\phi)$ and $g_*$ is the number of SM degrees of freedom.
To calculate $\Gamma_\varphi$, we first calculate the mass of the inflaton $m_\varphi$ from the numerically calculated second derivative of the RGE improved potential at the minima $v_{BL}/\sqrt{2}$,  $V''(\phi)|_{\phi=v_{BL}/\sqrt{2}}=m_\phi^2=2m_\varphi^2$. For our benchmark point of $M=M_P$ and $\mu_{T}=44.85$ TeV, we have $v_{BL}=49.1$ TeV as shown in Fig. \ref{fig:minphi}, and $m_\varphi=0.85$ GeV. 

We now note that the SM Higgs boson mass ($m_h = 125$ GeV) is given by, 
\begin{equation}
  m_h^2  = \lambda_{H\phi} v_{BL}^2, 
\label{eq:mH}
\end{equation} 
where $v_H = 246$ GeV is the Higgs doublet VEV.

The mass matrix of the Higgs bosons, $\varphi$ and $h$, is given by,
\begin{eqnarray}
{\cal L}  \supset -
\frac{1}{2}
\begin{bmatrix}
h & \varphi
\end{bmatrix}
\begin{bmatrix} 
m_{h}^2 &  \lambda_{H\phi} v_{BL} v_{H} \\ 
 \lambda_{H\phi} v_{BL} v_{H} & m_{\varphi}^2
\end{bmatrix} 
\begin{bmatrix} 
h \\ \varphi 
\end{bmatrix}.  
\label{eq: massmatrix}
\end{eqnarray} 

Diagonalising the mass matrix by 
\begin{eqnarray}
\begin{bmatrix} 
h \\ \varphi 
\end{bmatrix}
=
\begin{bmatrix} 
\cos\theta &   \sin\theta \\ 
-\sin\theta & \cos\theta  
\end{bmatrix} 
\begin{bmatrix} 
{\tilde h} \\ {\tilde \varphi} 
\end{bmatrix}  ,
\label{eq: eigenstate}
\end{eqnarray} 
where ${\tilde h}$ and ${\tilde \varphi}$ are the mass eigenstates, and 
   the mixing angle $\theta$ determined by 
\bea
2 v_{BL} v_{H}  \lambda_{H\phi}= ( m_h^2 -m_\varphi^2) \tan2\theta,  
\label{eq: mixings} 
\eea  
we find, for $m_\varphi^2  \ll m_h^2$ and $\lambda_{H\phi} \ll1$,
\bea
\theta \simeq \frac{v_H}{v_{BL}}.
\label{eq:theta}
\eea

The mass eigenvalues are then given by
\bea
&&m_{\tilde{\varphi}}^2 = m_{\varphi}^2  + \left(m_\varphi^2  - m_h^2 \right) \frac{\sin^2\theta}{1-2 \sin^2\theta} 
  \simeq m_{\varphi}^2   - m_h^2 \theta^2,   \nonumber\\
&&m_{\tilde{h}}^2 = m_h^2 - \left(m_\varphi^2  - m_h^2 \right) \frac{\sin^2\theta}{1-2 \sin^2\theta} \simeq m_h^2. 
\label{eq: masses} 
\eea  
For the parameter values we are interested in, 
we find $m_{\tilde{\varphi}, {\tilde{h}}}  \simeq m_{\varphi, h}$ and $\tilde{\phi}, \tilde{h} \simeq \phi, h$.  We noted that, for this mass eigenstate approximations to make sense numerically, i.e. to really get $m_{\tilde{\varphi}, {\tilde{h}}}  \simeq m_{\varphi, h}$, we require approximately $m_{Z_{BL}}\gtrsim850 ~\rm GeV$\footnote{This approximate relation comes from the fact that we require $m_{\varphi}^2 >m_h^2 \theta^2=m_h^2\left(\frac{v_H}{v_{BL}}\right)^2$ to get $m_{\tilde{\varphi}}  \simeq m_{\varphi}$. We observed that, keeping $m_{Z_{BL}}=2g_{BL}v_{BL}>850 \rm GeV$ resolves this issue by having to large value of $v_{BL}$ in the denominator of $\theta$, hence making $\theta$ small enough for theoretical consistency.}.
So, for notational simplicity, we will refer to the mass eigenstates without using {\it tilde} in the rest of this work. 

Coming back to our benchmark point of $M=M_P$ and $\mu_{T}=44.85$ TeV, the inflaton decays into SM particles through mixing with SM Higgs, with the mixing angle  $\theta\simeq\frac{v_H}{v_{BL}}=\frac{246}{\sqrt{2}~49.1\times10^3}$. The dominant decay channel of the inflaton with mass $m_\varphi=0.85$ GeV is into strange quark or muon pairs. This decay rate of the inflaton is then given by, 
\be
\Gamma_\varphi=\Gamma_H \sin^2 \theta \simeq \left(\frac{3}{8\pi}\frac{m_s^2}{v_H^2}m_\varphi+\frac{1}{8\pi}\frac{m_\mu^2}{v_H^2}m_\varphi\right)\sin^2\theta,
\ee
where $\Gamma_H$ is the SM Higgs decay rate into pairs of SM particles with mass $<m_\varphi/2$ and $m_s\sim96$ MeV, $m_\mu\sim105$ MeV denotes mass of strange quark and muon particles respectively. This decay leads to the reheating temperature of 
$T_{rh}\simeq635~ \rm GeV.$

For the benchmark points $M=5.67$ and $0.1~M_P$, respectively $\mu_T=48.53$ and 
$96.63$ TeV, we get $v_{BL}=52.73$ and 
$106.06$ TeV, $m_\phi=2.93$ GeV (dominant decay channel into charm quark pair) and 
$0.39$ GeV (dominant decay channel into strange quark and muon pairs). Following the same prescription given in this section earlier, we get $T_{rh}\simeq12.3$ TeV and 
$198.02$ GeV respectively. So, all three benchmark points have $T_{rh}>1$ MeV (scale of BBN).


It is important to mention that to calculate a more realistic reheating temperature we need to solve the field dynamics after inflation in $\phi-H$ plane. This is because whenever the field trajectory has a component in the SM Higgs direction, which is indeed the case near the minima of the potential, we get a sudden suppression in the energy density in the fields due to high decay rate of the SM Higgs, hence helping the reheating cause. However calculating this dynamics is complex in our case due to sudden sharp edges of the potential near the threshold scale $\mu_T$, and we omit this analysis in this work. Keeping this in mind, we state that the $T_{rh}$ estimates we have done are more conservative, hence having this conservative estimates well over the BBN scale is enough for our bench mark points to be consistent with the standard Big Bang Cosmology.

We also mention that the reheating temperature $T_{rh}$ is far smaller than the inflationary Hubble scale,
$\mathcal{H}_{inf}=\frac{1}{\sqrt{3}M_P}\sqrt{V(\phi)|_{\phi=M}}$. 
For the benchmark points (M = 5.67 $M_P$, $\mu_T = $48.53 TeV), (M = 1 $M_P$, $\mu_T = 44.85$ TeV) and ($M = 0.1 ~M_P$, $\mu_T = 96.63$ TeV), scales of inflation are $\mathcal{H}_{inf}=2.79\times10^{12}$ GeV, $1.53\times10^{10}$ GeV and $1.53\times10^7$ GeV respectively, in contrast to $T_{rh}$ in the TeV scale. 

\section{Model Constraints from LHC}
\label{signatures}

In this section we will discuss the constraints on our model parameters that come from the gauge boson search in hadron colliders, via the s-channel process $p+p\to Z' \to e^+e^-/\mu^+\mu^-$.

For computing cross-section for LHC processes, $g_{BL}$ is already constrained to be small, and we interpret the $Z'$ search to be equivalent to $Z_{BL}$ search. The cross-section for the process $\sigma(pp\to Z_{BL}\to e^+e^-/\mu^+\mu^-)\simeq\sigma(pp\to Z_{BL}){\rm BR}(Z_{BL}\to e^+e^-/\mu^+\mu^-)$ in the narrow-width approximation,
\begin{align}
  \sigma(pp\to Z_{BL})=2\sum_{q,\overline q}\int dx\int dy\, f_q(x,Q)f_{\overline q}(y,Q)\hat\sigma(\hat s),
\end{align}
with
\begin{align}
  \hat\sigma(\hat s)=\frac{4\pi^2}{3}\frac{\Gamma(Z_{BL}\to q\overline q)}{M_{Z_{BL}}}\delta(\hat s-M_{Z'}^2).
\end{align}
where $f_q$ and $f_{\overline q}$ represent the parton distribution functions for a quark and an anti-quark, $\hat s\equiv xys$ represent the invariant squared mass of the quarks in collision. In our LHC Run-2 analysis we will follow Ref. \cite{Aad:2019fac}
which is for the c.o.m $\sqrt s=13\;{\rm TeV}$.

\begin{figure}[H]
\begin{center}
\includegraphics[height=0.43\textwidth]{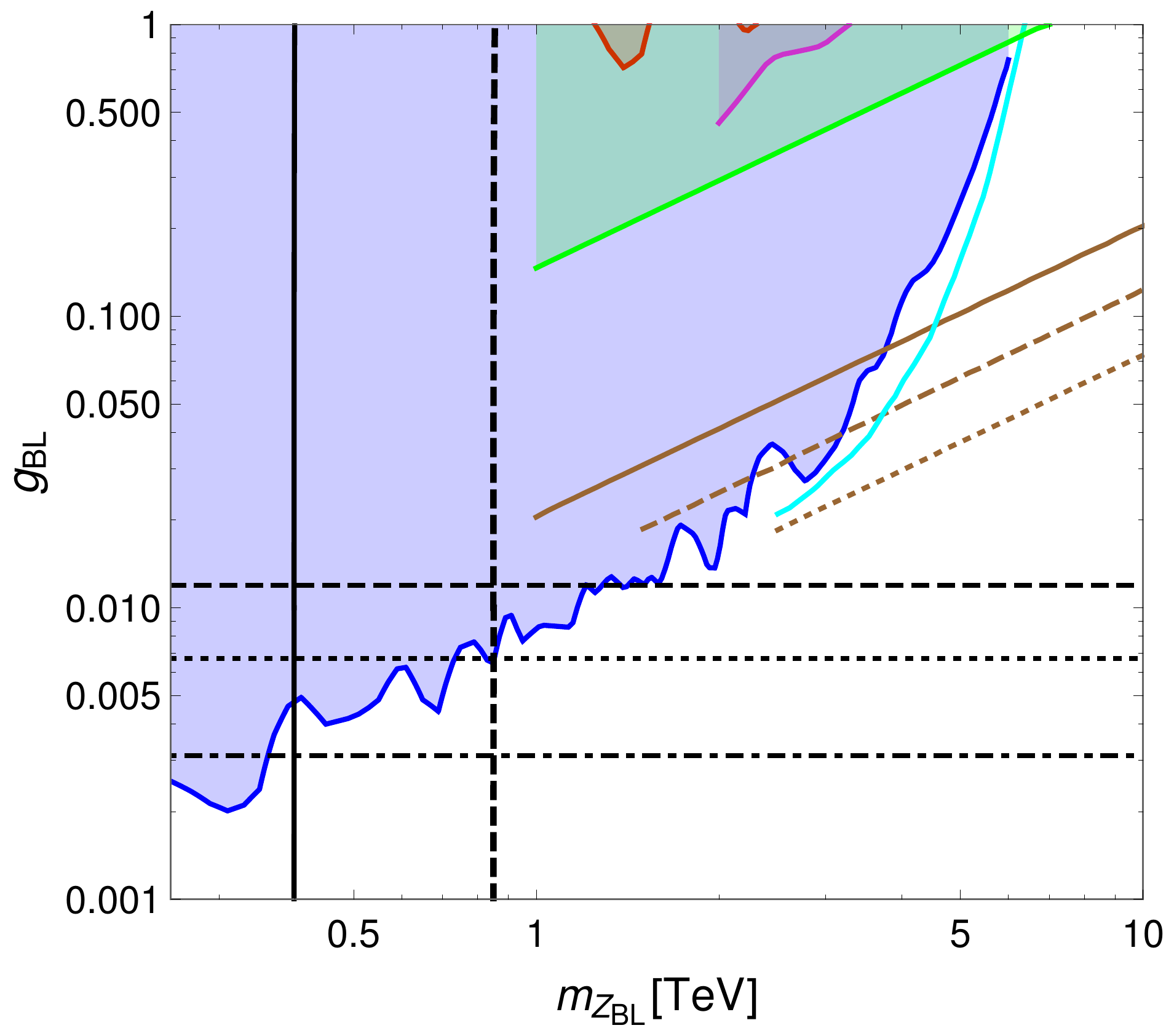}
\includegraphics[height=0.43\textwidth]{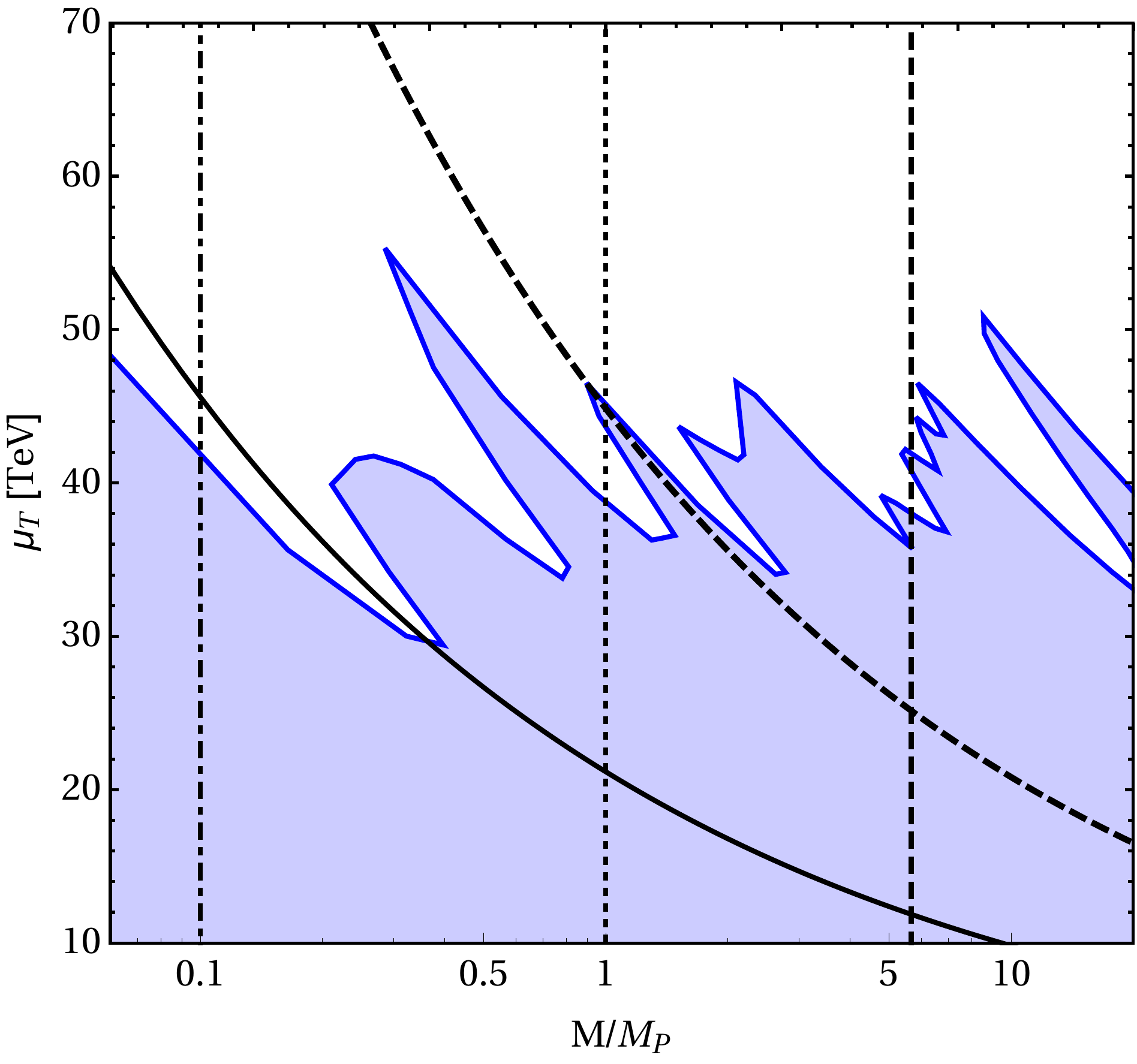}
	\caption{\it \textbf{Left Panel:} The horizontal lines correspond to the inflection-point scale M = 5.67 $M_P$ (dashed), $M_P$ (dotted) and 0.1 $M_P$ (dot-dashed) respectively \footnote{Theoretical upper bound from validity of the inflationary calculations as per Eqn(\ref{upper_bound}).}. This corresponds to $m_{Z_{BL}}$ lower bounds to be 1.64 TeV, 850 GeV and 360 GeV respectively, when taken into account $B-L$ gauge coupling limit from ATLAS di-lepton result, as mentioned later. The vertical solid line and the vertical thick dashed line correspond to $m_{Z_{BL}}=3\times133$ and $850$ GeV respectively, the lower limit for theoretical consistency as discussed in the text.  The jagged blue solid line 
	is the upper bound on the $B-L$ gauge coupling
 as a function of $Z_{BL}$ mass from the ATLAS dileption final result (ATLAS-CONF-2019-001) \cite{ATLAS:2019vcr}. We also explicitly show the other current constraints (dashed region): Green corresponds to LEP-II \cite{ALEPH:2013dgf}, Magenta and Red are bounds obtained from ATLAS13 (2j)\cite{ATLAS:2019vrv}, CMS13 (2j)\cite{CMS:2018mgb}. Cyan ATLAS-TDR (2l)\cite{CERN-LHCC-2017-018}. Brown solid, dashed, dotted are for future e+e- colliders with $\sqrt{s}=$ 250, 500 and 1000 GeV. The LEP and the future e+e- collider bounds were obtained using contact interaction strategy in the Refs.\cite{Das:2021esm,Das:2022oyx}.
 \textbf{Right Panel:} Plot of $M/M_P$ versus $\mu_T$ (TeV). Using the relation $m_{Z_{BL}}= 2 g_{BL} v_{BL} \sim 2\sqrt{2} g_{BL} \mu_T$, it has been mapped to $M/M_P$ vs $\mu_T$ parameter space via Eq.\ref{FEq2}. The open parameter region bound from right by $M=5.67~M_P$, from bottom by the ATLAS final result and from left by the thick dashed curve corresponding to $m_{Z_{BL}}=850$ GeV is allowed within our theoretical framework and experimental constraints.  
 }
\label{fig:atlas}
\end{center}
\end{figure}

Current experimental constraints in the $m_{Z_{BL}}$ vs $g_{BL}$ parameter space corresponding to $B-L$ vector boson (as shown in left panel of Fig. \ref{fig:atlas}), with mass $m_{Z_{BL}}= 2 g_{BL} v_{BL}$, can be mapped to $M/M_P$ vs $\mu_T$ parameter space\footnote{Note due to the abrupt change in value of $\lambda_\phi$, as shown in right panel of Fig. \ref{fig:rgeall}, we may use $\mu_T\sim v_{BL}$.} of our model using Eq.\ref{FEq2} and the expression $m_{Z_{BL}}= 2 g_{BL} v_{BL} \sim 2\sqrt{2} g_{BL} \mu_T$. Constraints from these experiments mentioned before into $M/M_P$ vs $\mu_T$ parameter space is shown in the right panel of Fig. \ref{fig:atlas}.

As one can see from Fig. \ref{fig:rgeall}, g$_{BL}$ and $y$ and $Y_{low}$ remains almost constant throughout the RGE evolution.
When the $U(1)_{B-L}$ symmetry is broken, the $Z'$ boson becomes massive with its $M_{Z'}=2g_{BL} v_{BL}$. Fig. \ref{fig:atlas} benchmark points (M = 0.1 Mp, M = Mp, 5.67 Mp) on the $m_{Z_{BL}}$-$g_{BL}(M_{\rm P})$ plane, computed at the $U(1)_{B-L}$ breaking scale $\mu = \phi=v_{BL}/\sqrt{2}$. The plots clearly indicate for $g_{BL} \sim O(10^{-2})$ the mass $Z_{BL}$ boson can be of $\mathcal{O}(TeV)$ to satisfy all constraints.  For lower $g_{BL}$ values even lighter $Z_{BL}$ mass is allowed, upto 
$\sim850$ GeV.

 
In terms of the free parameters of the model, namely $\mu_T$ and M, it can be seen from right panel of Fig. \ref{fig:atlas} that $\mu_T \lesssim 50$ TeV is excluded by LHC run-2 data. Note that the analysis is only consistent within the open region limited by $M<5.67~M_P$ and the curve corresponding to the condition $m_{Z_{BL}}>850$ GeV.

\bigskip

\section{Discussion \& Conclusion}\label{conclusion}


We investigated in a minimal $B-L$ conformal extension of the SM the possibility of the $B-L$ Higgs driving inflation in the UV (without any coupling to gravity). In this model, starting from the UV, we explicitly derived the conditions in order to mimic the SM Higgs mass generating mechanism dynamically via perturbative quantum corrections, \textit{\'a l\'a} Coleman-Weinberg in the IR. The main findings of the paper are as follows:
\begin{itemize}
    \item Cosmic inflation happens due to the flatness of the U$(1)_{B-L}$ Higgs potential achieved through bosonic and fermionic quantum corrections. Once the \textit{inflection-point scale M} is fixed, the values of the free parameters of the model, namely, $\lambda_\phi,~g_{BL},~y$ are fixed at the scale M, and its running via RGE determines its value at the lower EW scale.
    \item Due the nature of the running of $\lambda_{\phi}$ near the threshold correction energy scale $\mu_T$, the seesaw scale $v_{BL}/\sqrt{2} \sim \mu_T$ is a good approximation (see Figs. \ref{fig:rgeall} and \ref{fig:minphi}).
    \item Besides M, the only free parameter in the model is the threshold energy scale $\mu_T$ ($\sim v_{BL}/\sqrt{2}$), i.e. the scale at which $U(1)_{B-L}$ symmetry is broken. Once the symmetry is broken happens, $\phi$ obtains VEV $v_{BL}$ and a term similar to $-m^2|H|^2$ comes into play, mimicking the SM Higgs mechanism \textit{\'a l\'a} Coleman-Weinberg. 
    The condition for the SM Higgs mass generation determines the combination $ \lambda_{H \phi} \mu_T^2$ as $m_h^2 = \lambda_{H \phi} v_{BL}^2\sim2 \lambda_{H \phi} \mu_T^2$. Therefore, $\lambda_{H \phi}$ is fixed from Higgs VEV (246 GeV), once $\mu_T$ is fixed.
    \item Considering $Z_{BL}$ searches in LHC, particularly upper limits from ATLAS, we get $\mu_T \sim v_{BL}/\sqrt{2} \leq O(80)$ TeV is excluded (see Fig. \ref{fig:atlas}). 
    \item The model predicts $m_{Z_{BL}} \geq$ 2 TeV with $g_{BL} \sim O(10 ^{-2})$ to be consistent with the inflationary cosmology, dynamical generation of EW and Seesaw scales as well as allowed by LHC searches. Such a region will be within the reach of future experiments. 
    \item For the benchmark points (M = 5.67 $M_P$, $\mu_T = $48.53 TeV), (M = 1 $M_P$, $\mu_T = 44.85$ TeV) and ($M = 0.1 ~M_P$, $\mu_T = 96.63$ TeV) considered in the model we estimated the reheating temperature $T_{rh}\simeq$ 12.3 TeV, 635 GeV and 198.0 GeV respectively and the scales of inflation to be $\mathcal{H}_{inf}=2.79\times10^{12}$ GeV, $1.53\times10^{10}$ GeV and $1.53\times10^7$ GeV respectively,
    thus being consistent with BBN limits.
\end{itemize}
Conformal invariance dictates no scales are fundamental in nature. We showed here that the dynamical scale generation of electroweak physics (EW), heavy neutrino physics (seesaw scale) and the phenomena of inflation can be achieved together purely via quantum corrections in particle theory of fundamental interactions\footnote{Similar work was considered but in context to non-minimally coupled scalars in Ref. \cite{Kubo:2020fdd}.} and put constraints and predicted signatures in collider physics. To derive the Coleman-Weinberg potential, we only considered 1-loop fluctuations due to the bosonic and fermionic degrees of freedom. Since our model is minimally coupled to gravity, we do not expect the picture to change due to quantum corrections from the gravity sector. Finally, this work can be extended by a full two-field study of inflation and its effect on primordial non-Gaussianities (see, e.g. Ref. \cite{Wands:2007bd} for a review) or other observables like Primordial Blackholes (PBH) (see e.g. Ref. \cite{Garcia-Bellido:2017mdw}) and secondary Gravitational Waves predictions (see e.g. Ref. \cite{Domenech:2021ztg}). 

We envisage that our studies concerning the conformal invariance at the classical level, the interplay of dynamical mass scale generation in the IR and cosmic inflation in UV will open up a new direction in future to unified model-buildings in BSM theories, having to explain the dark matter, matter-antimatter asymmetry, inflation, Strong CP and the EW-Planck scales hierarchy problems, under one umbrella, with or without gravity, and have testable laboratory, astrophysical or cosmic observable predictions.


\section{Acknowledgements}
A.G. and A.P. thanks Arindam Das for discussions. This work is supported in part by the United States, Department of Energy Grant No. DE-SC0012447 (N.O.).


\bibliographystyle{apsrev4-1}

\bibliography{ref}
\end{document}